\documentclass[12pt]{article}
\usepackage{mathrsfs}
\usepackage{amssymb}
\usepackage{amsmath}
\usepackage[amsmath,thmmarks]{ntheorem}
\usepackage{fullpage}

\usepackage{graphicx}
\usepackage{bm}
\usepackage{slashed}

\newcommand{\be}{\begin{eqnarray}}
\newcommand{\ee}{\end{eqnarray}}

\begin{document}

\begin{titlepage}

\makebox[6.5in][r]{\hfill ANL-HEP-PR-11-83, NUHEP-TH/12-01, PSI-PR-12-01}

\vskip1.40cm
\begin{center}
{\Large {\bf W physics at the LHC with FEWZ 2.1}} \vskip.5cm
\end{center}
\vskip0.2cm

\begin{center}
{\bf Ryan Gavin$^1$, Ye Li$^{2,3}$, Frank Petriello$^{2,3}$, and Seth Quackenbush$^3$}
\end{center}
\vskip 6pt
\begin{center}
{\it $^1$Paul Scherrer Institut, CH-5232 Villigen PSI, Switzerland} \\
{\it $^2$Department of Physics \& Astronomy, Northwestern University, Evanston, IL 60208, USA} \\
{\it $^3$High Energy Physics Division, Argonne National Laboratory, Argonne, IL 60439, USA} \\
\end{center}

\vglue 0.3truecm

\begin{abstract}
\vskip 3pt \noindent
We present an updated version of the FEWZ (\textbf{F}ully
\textbf{E}xclusive \textbf{W} and \textbf{Z} production) code for the
calculation of $W^{\pm}$  and $\gamma^{*}/Z$ production at
next-to-next-to-leading order in the strong coupling.  Several new
features and observables are introduced, and an
order-of-magnitude speed improvement over the performance of FEWZ 2.0
is demonstrated. New phenomenological results for $W^{\pm}$ production
and comparisons with LHC data are presented, and used to illustrate
the range of physics studies possible with the features of FEWZ 2.1.
We demonstrate with an example the importance of directly comparing fiducial-region
measurements with theoretical predictions, rather than first extrapolating
them to the full phase space.
\end{abstract}

\end{titlepage}
\newpage

\section{Introduction \label{intro}}

Production of $W$ bosons plays several important roles in
hadron-collider physics studies.  At the Tevatron, this
channel furnishes the single most precise measurement of the $W$-boson
mass.  The determination of the charge asymmetry between $W^+$ and $W^-$
production at both the Tevatron and the Large Hadron Collider (LHC)
provides important constraints on quark and anti-quark parton
distribution functions (PDFs).  High-energy production of $W$ bosons
acts as a background to searches for the new $W^{'}$ bosons expected
in many gauge extensions of the Standard Model, while production of
the $W$ in association with jets is a major background to searches for
supersymmetry and other theories beyond the Standard Model with missing-energy signatures.
Production of $W$ bosons is copious at the LHC, and the statistical errors have already been rendered completely negligible with the few 
inverse femtobarns of data taken.  Only systematic errors remain, arising from both experimental and theoretical sources.  In many of the relevant observables, the experimental systematic errors are already at the few-percent level or smaller.

The theoretical understanding of $W$ production has reached an advanced stage.  The inclusive $O(\alpha_S^2)$ QCD corrections to electroweak gauge boson production have been known for some time~\cite{Hamberg:1990np}.
Exclusive production, which is necessary for any realistic prediction or phenomenological study in a detector 
of finite acceptance, is technically challenging but has been achieved~\cite{Anastasiou:2003yy,Anastasiou:2003ds,Melnikov:2006di,Melnikov:2006kv,Catani:2009sm,Catani:2010en}.  The electroweak and QED corrections are known~\cite{Baur:1998kt,Dittmaier:2001ay}, including the leading-logarithmic terms arising 
from multiple-photon emission~\cite{CarloniCalame:2006zq}.  Initial
efforts toward the combination of higher-order QCD and electroweak
corrections have been
made~\cite{Cao:2004yy,Balossini:2009sa,Bernaciak:2012hj}, and even
progress toward the exact mixed QED-QCD~\cite{Kilgore:2011pa} and
next-to-next-to-leading order QED corrections~\cite{Boughezal:2011jf}
has occurred.  The remaining theoretical uncertainties are estimated to be at the percent level for the majority of relevant observables.  These corrections are available to the experimental community in a variety of simulation codes.

We have previously released a 2.0 version of the simulation code FEWZ ({\bf F}ully {\bf E}xclusive ${\bf W}$ and ${\bf Z}$ Production) that implements the NNLO QCD predictions for neutral-current $\gamma^*/Z$ production at hadron colliders, and allows for arbitrary kinematic cuts to be imposed~\cite{arXiv:1011.3540}.  FEWZ features a parallelized integration routine suited for running on modern computer clusters, allows for multiple, arbitrary kinematic variables to be binned during a single run, and automatically calculates PDF errors for the total cross section and all histogram bins.  It has been used by collaborations at the Tevatron, the LHC, and RHIC in their physics studies.  The purpose of this work is to apply the advances in FEWZ 2.0 to the description of $W$ production also, and to introduce further improvements to the FEWZ framework, resulting in a FEWZ 2.1 version.  We have further improved the integration routine to provide an additional order-of-magnitude gain in speed. FEWZ 2.1 incorporates the LHAPDF format~\cite{Whalley:2005nh} to allow all PDF sets of interest to be studied.  Several new observables are introduced in the new version of the program.

Our manuscript is organized as follows. We discuss the new features of
FEWZ 2.1 in Section~\ref{sec:features}.  Both the new available
observables and the performance increase are detailed.  In
Section~\ref{sec:numbers} we present phenomenological results for $W$
physics at the LHC.  We present comparisons of both integrated cross
sections within the fiducial region and the charge asymmetry with
ATLAS data.  Several ratios of $W^+$ over $W^-$ distributions are
shown, both to illustrate the range of studies possible with the scripts
distributed with FEWZ 2.1, and to highlight physics features of $W$
production at the LHC.  The importance of directly comparing
fiducial-region measurements to theoretical predictions, rather than
first extrapolating them to the full phase space, is demonstrated
using the $W^+$ over $W^-$ cross section ratio.  Only before the
extrapolation is the measurement error small enough to allow for
discrimination between different PDF sets.  We conclude in Section~\ref{sec:conc}.

\section{New features \label{sec:features}}

Version~2.1 applies all improvements of FEWZ 2.0~\cite{arXiv:1011.3540}
to $W$ boson production.  The user may now make either of two
executables, \texttt{fewzw} or \texttt{fewzz}, which correspond to $W$
or $\gamma^{*}/Z$ production, respectively.  Most source code is
shared between the two executables.  The sector structure is slightly
different owing to a few differing diagrams and symmetry factors at
NNLO between $W$ and $\gamma^{*}/Z$ production.  Both executables take similar basic input files, and the same
histogram input.  The user of FEWZ 2.0 will not notice significant
differences in compiling and running the new version of the program.
We highlight below several new features of FEWZ 2.1.

\subsection{New observables}

We have added two new observables to further facilitate physics
studies.  
\begin{itemize}
\item Beam thrust~\cite{arXiv:0910.0467} has been added as an
  available histogram in FEWZ 2.1.  Beam thrust is defined in FEWZ 2.1 as a sum over observed jets~\cite{arXiv:1005.4060}:
\be
\tau_B = \frac{1}{Q} \sum_k |\vec{p}_{k,T}|e^{-|\eta_k - Y|},
\ee
where $Q$ and $Y$ are the dilepton invariant mass and rapidity, and
$\vec{p}_{k,T}$ and $\eta_k$ refer to the transverse momentum and
pseudorapidity of the $k$-th jet.  Beam thrust permits a central jet
veto to be imposed without a jet algorithm, making it more easily
amenable to resummation techniques.  We include it to facilitate
comparison between fixed-order results and resummation results, as
well as for study in its own right.

\item The transverse mass has been added to allow for $W$ production
  studies.  Transverse mass is defined as
\be
M_T = \sqrt{2p_{Tl} E_{Tmiss} (1-\cos(\Delta \phi_{l,miss}))}.
\ee
$p_{Tl}$ refers to the transverse momentum of the charged lepton,
$E_{Tmiss}$ refers to the transverse momentum of the neutrino, and
$\Delta \phi_{l,miss}$ denotes the angular separation between them in
the transverse plane.  $M_T$ has a Jacobian peak sensitive to the mass of the $W$ without being dependent on the unobserved longitudinal momentum of the neutrino.  It is defined analogously for the $Z$ for comparison purposes.

\end{itemize}
Both new observables are available for creating histograms; cuts on
transverse mass are also accessible in the input file.

\subsection{New output features \label{output}}
Several new possibilities for controlling the histogram output have
been added.
\begin{itemize}

\item The user now has the option to output cumulative histograms in addition to traditional ones.  This can be useful to study the effect of cut placement on the variable of interest.  This feature adds more than the convenience of avoiding adding histogram bins by hand: the resulting technical precision on a cumulative bin is better due to cancellations between bins.

\item Most experimental analyses use bin sizes that vary over the range of the histogram.  Formerly this could be handled in FEWZ at the cost of multiple runs, but now the user has the option to input their own histogram bounds for any given histogram.

\item In addition to the scripts of FEWZ 2.0 which allow one to
  perform addition and multiplication operations on the output of
  separate runs, an asymmetry operator has been added which directly
  combines the results of $W^+$ and $W^-$ runs in the following way:
\be
A_W = \frac{\sigma_{W^+}-\sigma_{W^-}}{\sigma_{W^+}+\sigma_{W^-}}
\ee
Both the statistical and PDF errors are propagated consistently.

\end{itemize}

\subsection{Speed improvements}
The sector combinations detailed in~\cite{arXiv:1011.3540} have been
reworked to provide a modest reduction in variance of the integrand,
reducing the time required to reach the target precision.  Some of
these sectors have undergone ``sector recomposition," which patches together
the pieces resulting from sector
decomposition~\cite{hep-ph/0311311,hep-ph/0501130} by undoing the
variable transformations needed in the case of singularities in
multiple endpoints of the phase-space integrals.  This results in a
more consistent mapping of the integration variables to parton phase
space for similar sectors.  There are now 133 sectors for
\texttt{fewzz} and 154 for \texttt{fewzw}.  Alternative techniques for
reducing the number of required integrals when applying sector
decomposition have been discussed in Ref.~\cite{Anastasiou:2010pw}.  We have also been careful
to eliminate redundancies caused by calling the same code with the
same arguments repeatedly.  Instead, the results of these calls (where
they cannot be avoided) are stored in a cache or buffer.  Finally, we have rearranged how certain variables are stored, so that access is more likely to occur contiguously in memory, resulting in fewer CPU cache misses.  This can make a large difference in the case of parton distribution functions, which are accessed frequently, or histograms, which are filled frequently.

\begin{figure}[h!]
\begin{center}
\includegraphics[scale=0.4,angle=90]{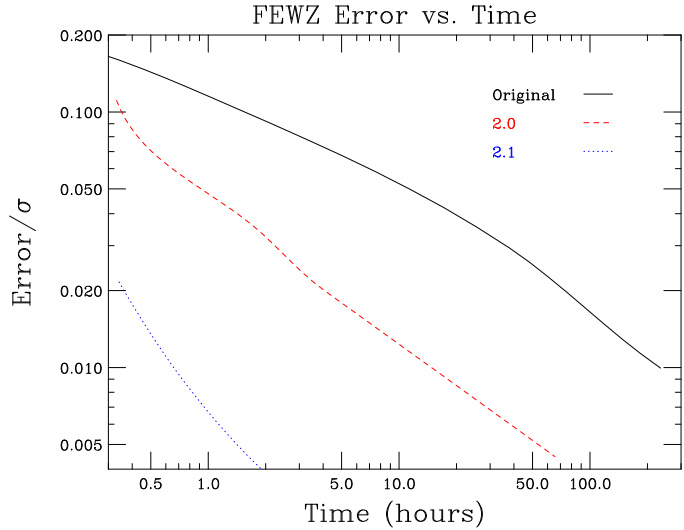}
\caption{Relative error versus time for different versions of FEWZ.}
\label{speedfig}
\end{center}
\end{figure}

Altogether, the speed improvements are significant.  In
Fig.~\ref{speedfig}, we reproduce Fig.~1 of~\cite{arXiv:1011.3540},
which shows integration precision as a function of time.   FEWZ 2.1
has been added to the comparison.  The cuts, histograms, PDF sets, and
benchmark machine are the same as Ref.~\cite{arXiv:1011.3540}, with
the exception of the addition of the two new histograms described in
Sec.~\ref{output}.  The time required to reach a target precision is
an order of magnitude less than needed for FEWZ 2.0.

\section{Example results \label{sec:numbers}}
To demonstrate the capabilities of FEWZ 2.1, we present in this section
several results relevant for the study of $W$-boson production at the
LHC.  We impose the following set of standard acceptance cuts on the
$W$-decay products:
\begin{eqnarray}
M_T > 40 \, \text{GeV}, && E_{Tmiss}> 25\, \text{GeV}, \nonumber \\
p_{Tl} > 20 \, \text{GeV}, && |\eta_l| < 2.5.
\label{cuts}
\end{eqnarray}
We begin by studying the total cross section for $W^{\pm}$ production
subject to these cuts, together with the ratio of $W^+$ over $W^-$
within this phase-space region.  To illustrate the sensitivity of the cross
section to different extractions of parton distribution functions, we
show results for three NNLO PDF fits: those of HERA~\cite{CooperSarkar:2011aa},
MSTW~\cite{Martin:2009iq}, and NNPDF~\cite{Ball:2011uy} (to avoid
clutter in the plots and tables we do not show the NNLO results of ABKM~\cite{Alekhin:2009vn} or JR~\cite{JimenezDelgado:2009tv},
which have been presented in our previous work on $Z$ physics~\cite{arXiv:1011.3540}).
\begin{eqnarray}
\text{HERA:}&& \sigma_{W^-}=2134^{+15}_{-22}\;
\text{pb};\;\; \sigma_{W^+}=3265^{+37}_{-31}\; \text{pb};
\;\; \frac{\sigma_{W^+}}{\sigma_{W^-}} = 1.530^{+0.017}_{-0.015}\nonumber \\
\text{MSTW:}&& \sigma_{W^-}=2106^{+35}_{-35}\; \text{pb};\;\; \sigma_{W^+}=3181^{+57}_{-55}\; \text{pb}; \;\; \frac{\sigma_{W^+}}{\sigma_{W^-}} = 1.510^{+0.014}_{-0.013}\nonumber \\
\text{NNPDF:}&& \sigma_{W^-}=2085^{+32}_{-32}\; \text{pb};\;\; \sigma_{W^+}=3255^{+52}_{-52}\; \text{pb};\;\; \frac{\sigma_{W^+}}{\sigma_{W^-}} = 1.563^{+0.026}_{-0.026}.
\end{eqnarray}
The 68\% C.L. PDF errors are shown for each set.  The parametric
uncertainties arising from imprecise knowledge of the strong coupling and other
parameters are not included in these results. Scale uncertainties have
previously been shown to be subdominant to PDF uncertainties for electroweak gauge boson
production at the LHC~\cite{arXiv:1011.3540,Watt:2011kp}.  We note that the technical
errors arising from the integration error are less than 0.1\%.  The
results for each PDF set are consistent, although the MSTW results for
$\sigma_{W^+}$ and the ratio are more than one sigma lower than the
results for the other sets.  We compare the results above with the
ATLAS results for the $\sigma_{W^+}/\sigma_{W^-}$
ratio~\cite{Aad:2011dm}:
\be
\left(\frac{\sigma_{W^+}}{\sigma_{W^-}}\right)_{\text{fid}} = 1.542 \pm 0.007\text{(stat.)} \pm
0.012\text{(sys.)} \pm 0.001\text{(accept.)}.
\label{atlasrat}
\ee
We note that ATLAS combines both the electron and muon channels and extrapolates them to a common fiducial region when obtaining this number. The subscript 'fid.'
denotes that the ratio is measured in this common region, which coincides exactly
with the phase space defined by our standard acceptance cuts. The last error
labeled ‘accept.’ denotes the uncertainty contribution arising from
extrapolation.  The NNPDF and HERA PDF predictions are consistent
within one sigma from the ATLAS measurement, while the MSTW result
differs by over two sigma from the central value of the
measurement.  The situation becomes murkier if the ratio of
inclusive cross sections is compared instead.  In that case, the
measured fiducial cross section is extrapolated to the full phase
space using Monte Carlo simulation, introducing an additional source of uncertainty.  The ATLAS
measurement of the inclusive ratio is
\be
\left(\frac{\sigma_{W^+}}{\sigma_{W^-}}\right)_{\text{inc}} = 1.454 \pm 0.006\text{(stat.)} \pm
0.012\text{(sys.)} \pm 0.022\text{(accept.)}.
\label{atlasinc}
\ee
Upon combining the various sources of uncertainty in quadrature, an
error twice is large than the one in Eq.~(\ref{atlasrat}) is
obtained.  This result should be compared with the theoretical
predictions for the inclusive cross section ratio:
\begin{eqnarray}
\text{HERA}: && \left(\sigma_{W^+}/ \sigma_{W^-}\right)_{\text{inc}} = 1.440^{+0.015}_{-0.014}; \nonumber \\
\text{NNPDF}: && \left(\sigma_{W^+}/\sigma_{W^-}\right)_{\text{inc}} = 1.471^{+0.020}_{-0.020}; \nonumber \\
\text{MSTW}: && \left(\sigma_{W^+}/\sigma_{W^-}\right)_{\text{inc}} = 1.435^{+0.013}_{-0.010}.
\end{eqnarray}
All three sets now agree within the estimated one-sigma uncertainty
with the results in Eq.~(\ref{atlasinc}).  The full potential of the
experimental measurement to distinguish between different PDF
extractions is only realized if the fiducial cross sections can be
directly compared to theoretical predictions.  With FEWZ 2.1, this
becomes possible.  We note that a comparison of CMS results~\cite{CMS:2011aa} with the $W^+$ over $W^-$ inclusive
cross section ratio obtained with different NNLO PDF sets was
presented in Ref.~\cite{Watt:2011kp}, where a similar point regarding
the importance of studying the fiducial cross sections was made.  In that study, parametric
uncertainties on the ratio of $W^+$ over $W^-$ cross sections were
also studied, and found to increase the PDF-only error by a negligible
amount.  We rely on this result to safely neglect them here.

\begin{figure}[h!]
\begin{minipage}[b]{3.5in}
  \includegraphics[width=3.0in]{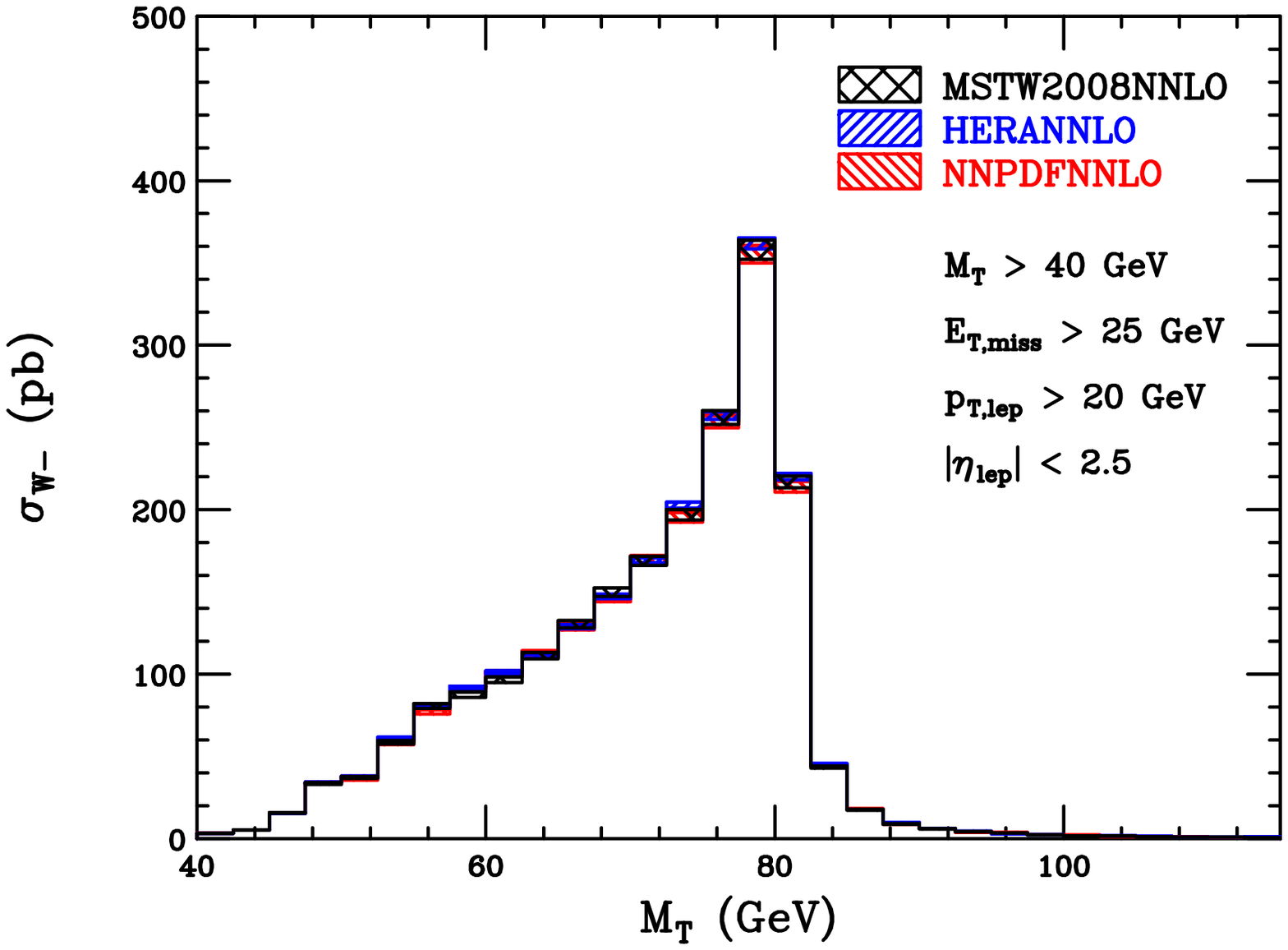}
\end{minipage}
\begin{minipage}[b]{3.5in}
  \includegraphics[width=3.0in]{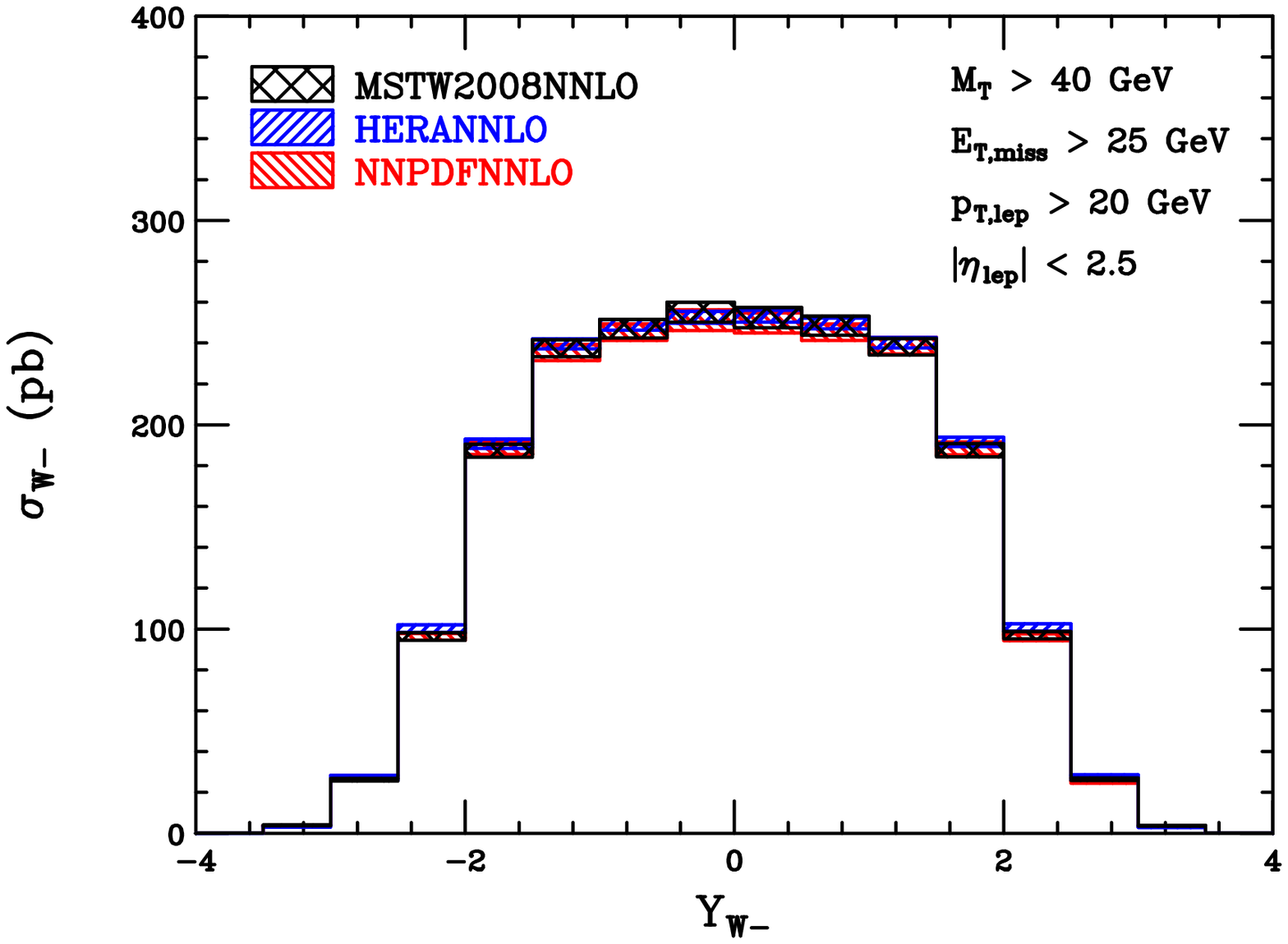}
\end{minipage}
\\
\begin{minipage}[b]{3.5in}
\vspace{1cm}
  \includegraphics[width=3.0in]{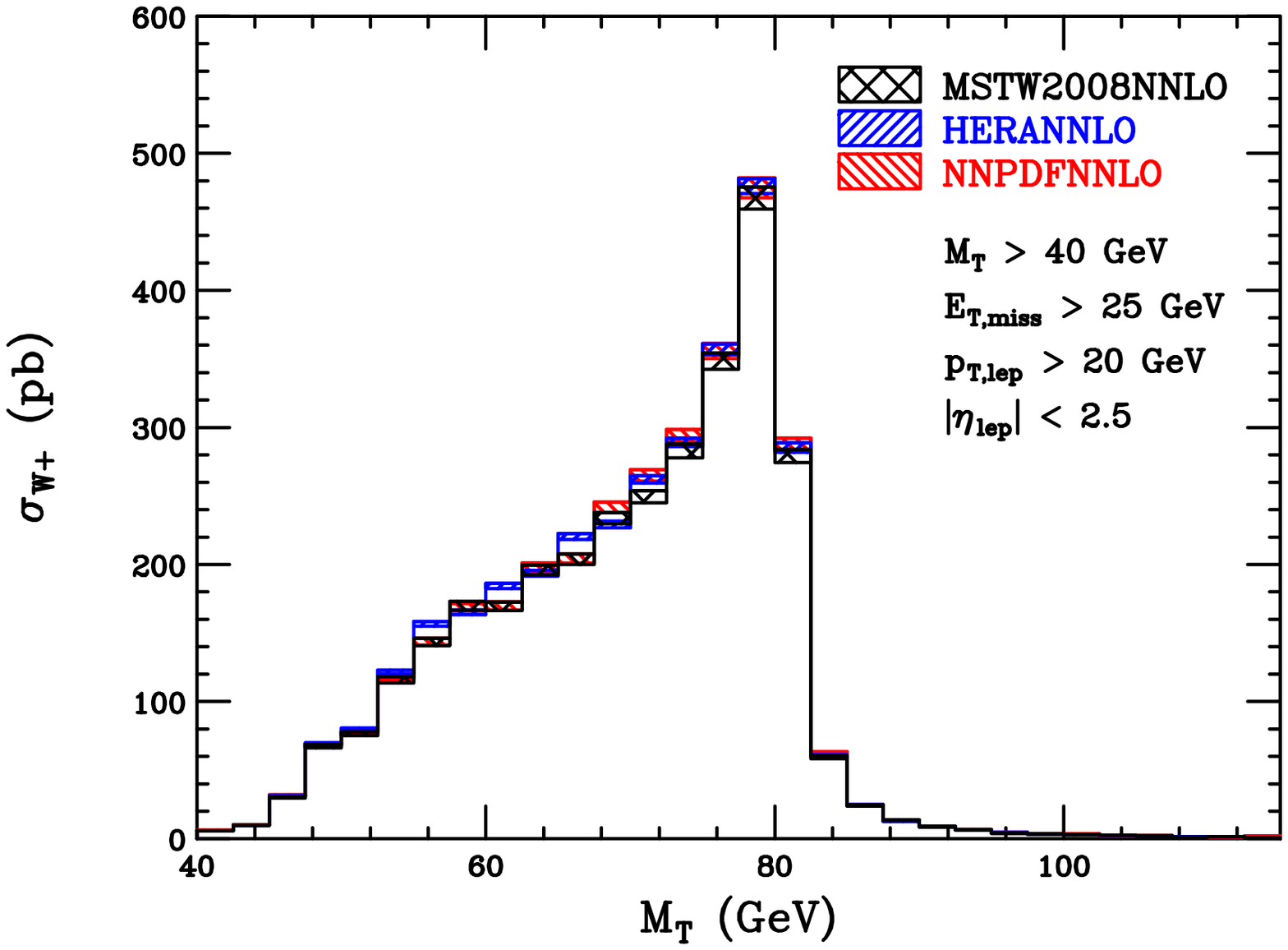}
\end{minipage}
\begin{minipage}[b]{3.5in}
\vspace{1cm}
  \includegraphics[width=3.0in]{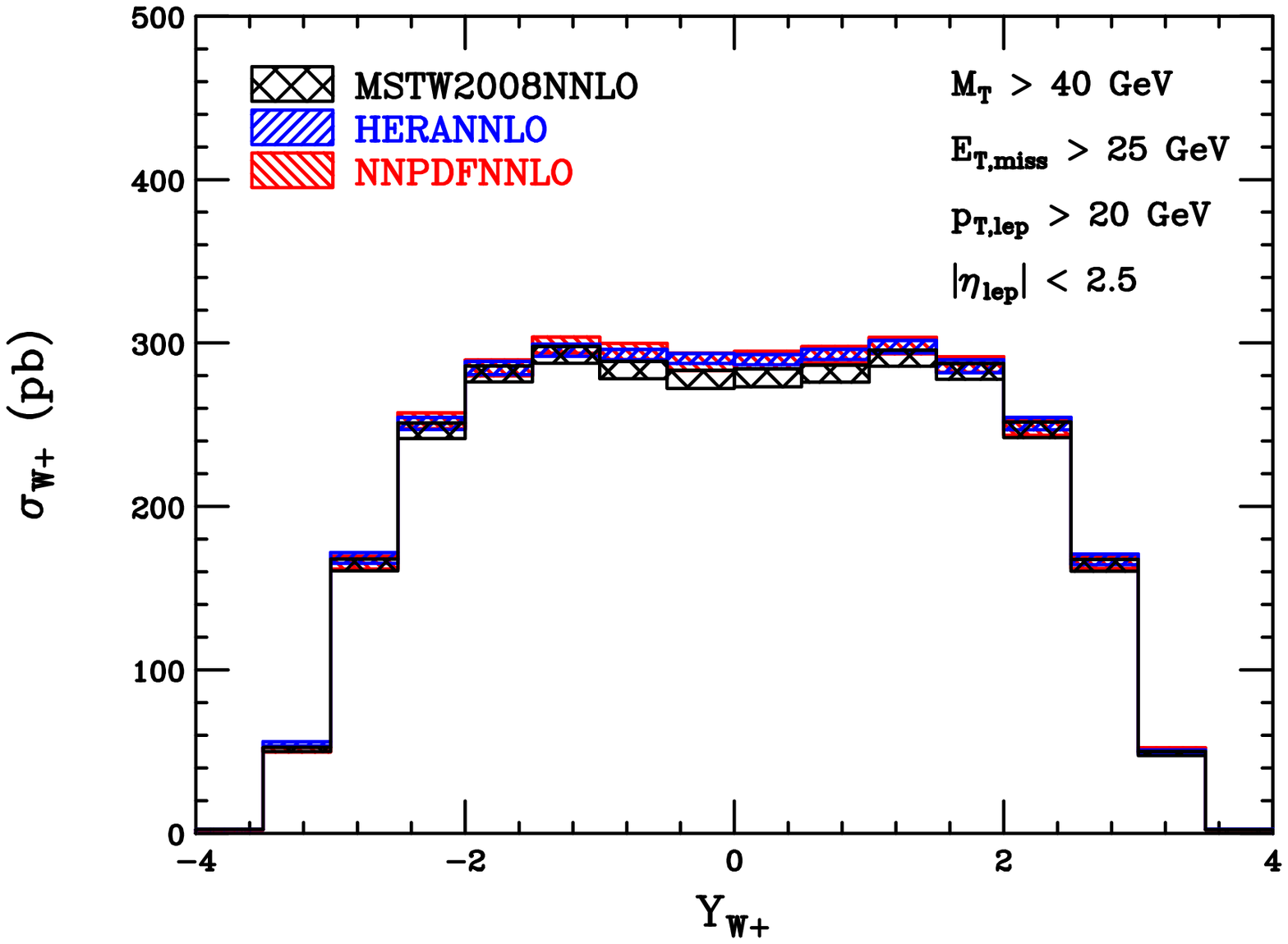}
\end{minipage}
\caption{Bin-integrated cross sections for the $W^-$ transverse mass
  (upper left panel), $W^-$rapidity (upper right panel), $W^+$ transverse mass
  (lower left panel), and $W^+$rapidity (lower right panel)  for all three NNLO PDF sets.  
The standard acceptance cuts of Eq.~(\ref{cuts}) have been implemented.  The bands indicate the PDF uncertainties for each set.
\label{basic}}
\end{figure}

We next present results for several distributions, to illustrate both the
functioning of FEWZ 2.1 and aspects of $W$ physics at the LHC.  The transverse
mass and rapidity distributions for $W^{\pm}$ are shown in
Fig.~\ref{basic}.  The Jacobian peaks for $M_T=M_W$ are clearly seen
in both plots.  Also visible is a similar, although less prominent, feature at $M_T = 50$ GeV
caused by the cuts of Eq.~(\ref{cuts}).  At leading order, the missing
$E_T$ and the lepton momentum must be equal and back-to-back in the
transverse plane, leading to the noted minimum.  This constraint is
relaxed at higher orders when the $W^{\pm}$ can recoil against
additional radiation, populating the region $40 \, \text{GeV} < M_T <
50$ GeV.  The stronger peaking of the up-quark PDF as compared to the
down-quark distribution at $x \sim 0.1$ is visible in the enhancement
of the $W^+$ rapidity distribution near $|Y| \sim 2$.

\begin{figure}[h!]
\begin{minipage}[b]{3.5in}
  \includegraphics[width=3.0in]{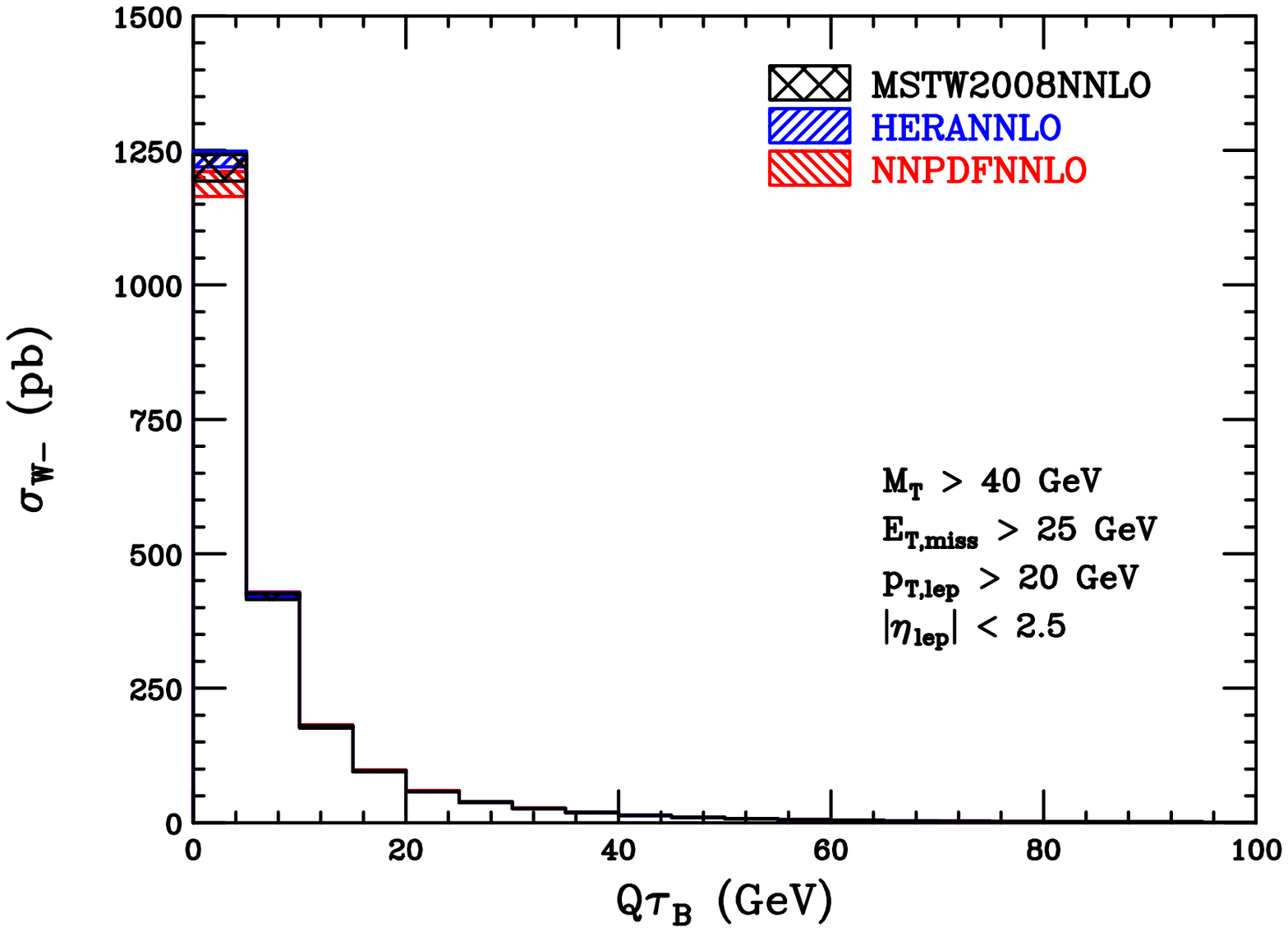}
\end{minipage}
\begin{minipage}[b]{3.5in}
  \includegraphics[width=3.0in]{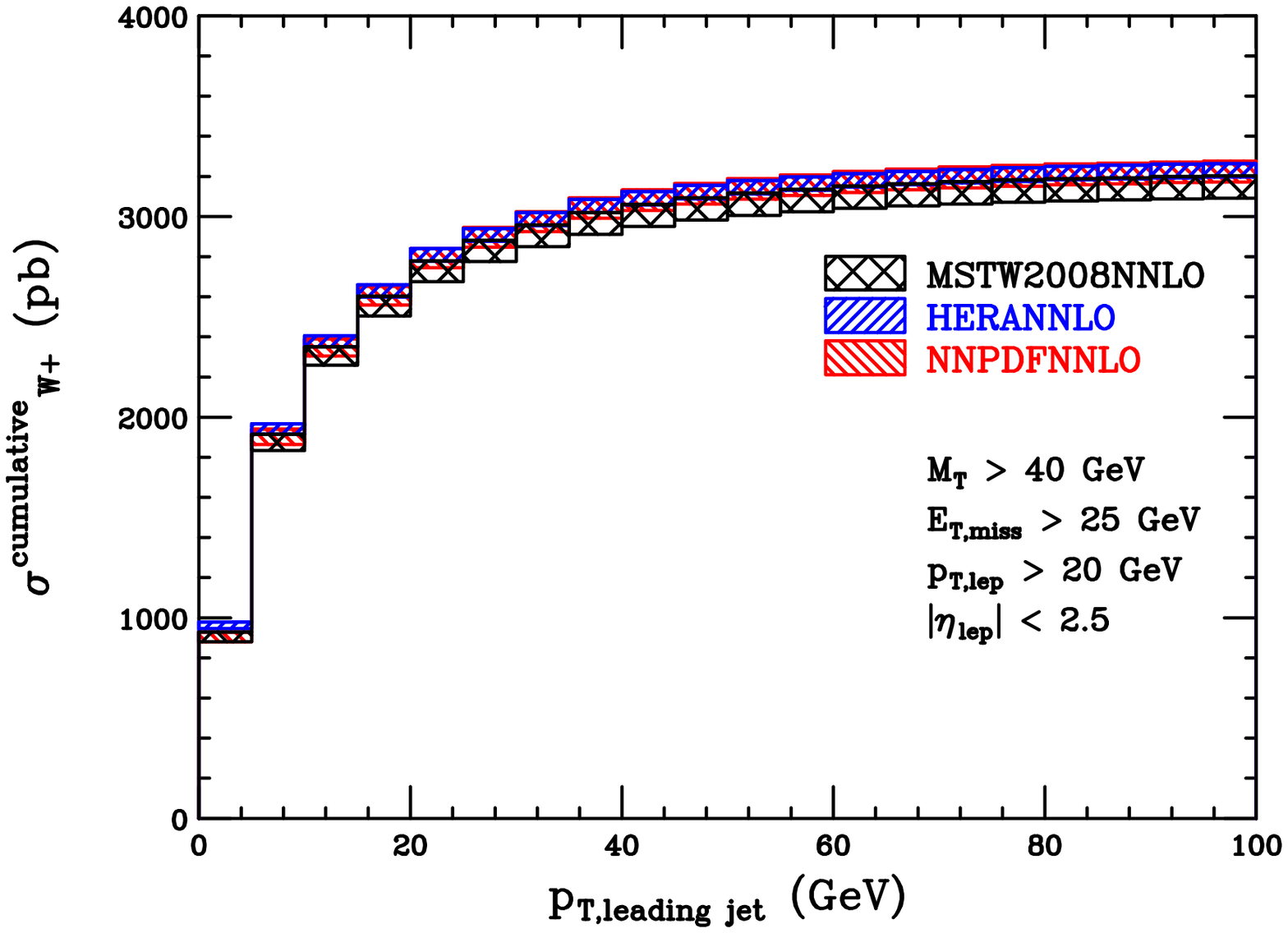}
\end{minipage}
\caption{Bin-integrated cross sections for the $W^-$ beam-thrust
  distribution (left panel) and the cumulative leading-jet $p_T$
  distribution in $W^+$ production (right panel).}
\label{new}
\end{figure}

In order to illustrate new features present in FEWZ 2.1, we present in
Fig.~\ref{new} results for the $W^-$ beam thrust and the
$W^+$ cumulative leading-jet $p_T$ distributions.  No significant
distributions between the three PDF sets are found for either observable.  Additional
radiation in $W$ production goes like $C_F \alpha_s /\pi$.  This is a
smaller quantity than the $C_A \alpha_s /\pi$ relevant for
gluon-initiated processes, and explains the peaking of
the beam-thrust distribution at lower values than observed in
Ref.~\cite{Berger:2010xi} for Higgs production and the quick plateau
observed in the leading-jet transverse momentum distribution.

The scripts distributed with FEWZ allow ratios of kinematic
distributions to be easily studied by combining the results of
separate runs.  We present below in
Fig.~\ref{ratios} several examples relevant to LHC phenomenology.  The
ratio $W^+$ over $W^-$ production as a function of both lepton
pseudorapidity and transverse momentum is shown in Fig.~\ref{ratios}
for the MSTW, NNPDF, and HERAPDF PDF sets.  
In the left panel, a separation between the sets is apparent at central rapidity, larger than the
estimated PDF uncertainty.  The ratio of lepton $p_T$ distributions
shown in the right panel reveals a harder distribution for the $W^-$
than the $W^+$.  This can be understood by considering the
leading-order kinematics.  The $p_T$ and $\eta$ of the lepton can be
written in terms of the center-of-momentum frame scattering angle as
$p_T = \sqrt{\hat{s}} \,\text{sin}\, \theta\, /2$, $\eta = -\text{ln
  tan} \,(\theta/2)$.  The larger fraction of events at large rapidity
for $W^+$ observed in Fig.~\ref{basic} translates into a larger
fraction of events at larger lepton pseudorapidity for $W^+$ than for $W^-$.  The leading-order kinematics
implies that $W^+$ events with large $\eta$ have scattering angles with
small $\text{sin}\,\theta$, and consequently smaller lepton $p_T$ than those
for $W^-$ production.

The
$W$ charge asymmetry as a function of lepton pseudorapidity, defined as 
\be
\frac{dA_W}{d\eta_l} = \frac{d\sigma_{W^+}/d\eta_l - d\sigma_{W^-}/d\eta_l}{d\sigma_{W^+}/d\eta_l + d\sigma_{W^-}/d\eta_l}
\ee
is well-known to provide a strong constraint on PDFs.  We compare in
Fig.~\ref{charge} the bin-integrated results for $A_W$ for the MSTW,
NNPDF, and HERAPDF sets with recent results from
ATLAS~\cite{Aad:2011dm}.  The HERAPDF and NNPDF predictions are in
good agreement with the data over the experimentally available range
of pseudorapidity, while the MSTW predictions are slightly lower
throughout this range. 

\begin{figure}[h!]
\begin{minipage}[b]{3.5in}
  \includegraphics[width=3.0in]{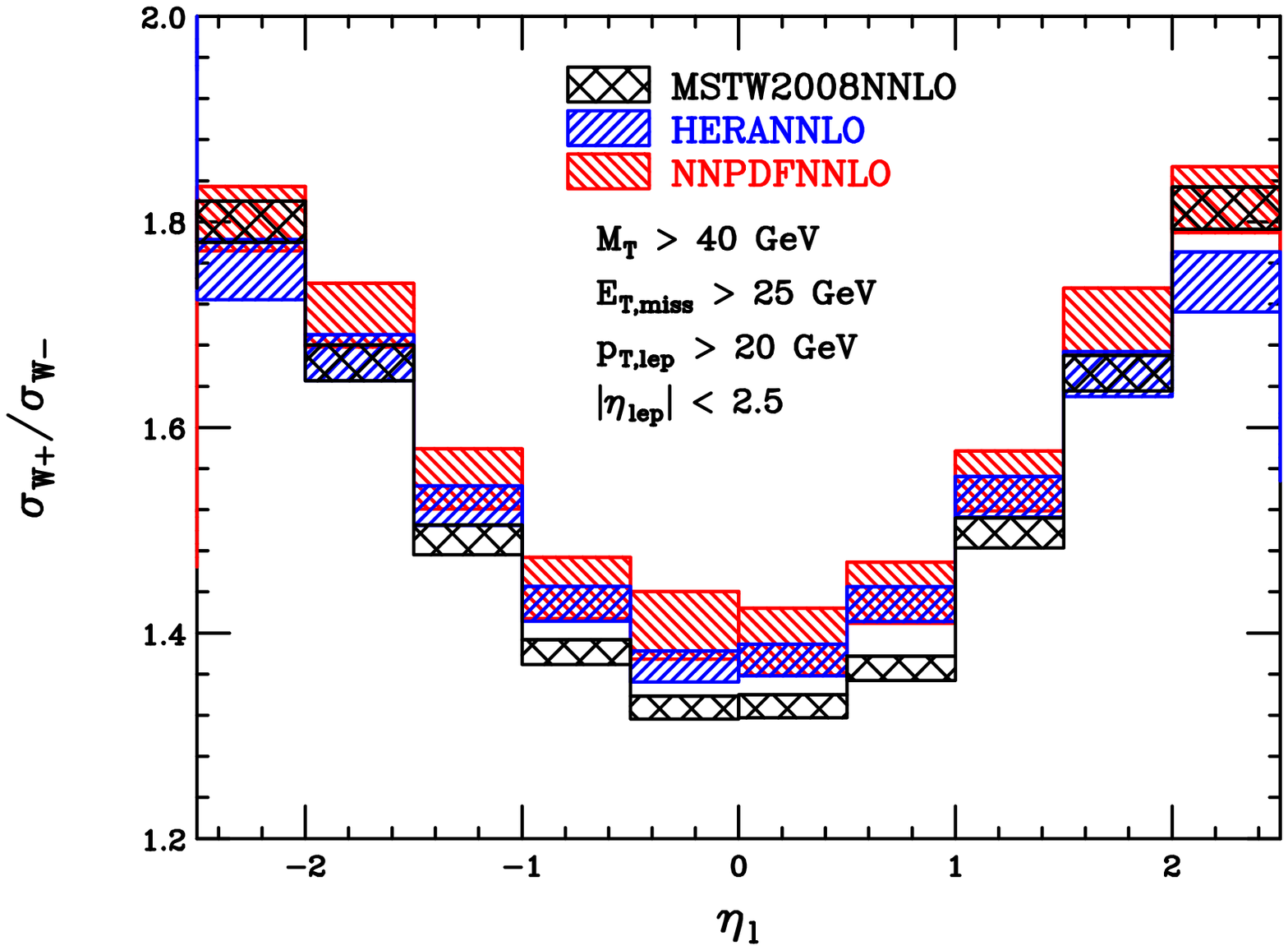}
\end{minipage}
\begin{minipage}[b]{3.5in}
  \includegraphics[width=3.0in]{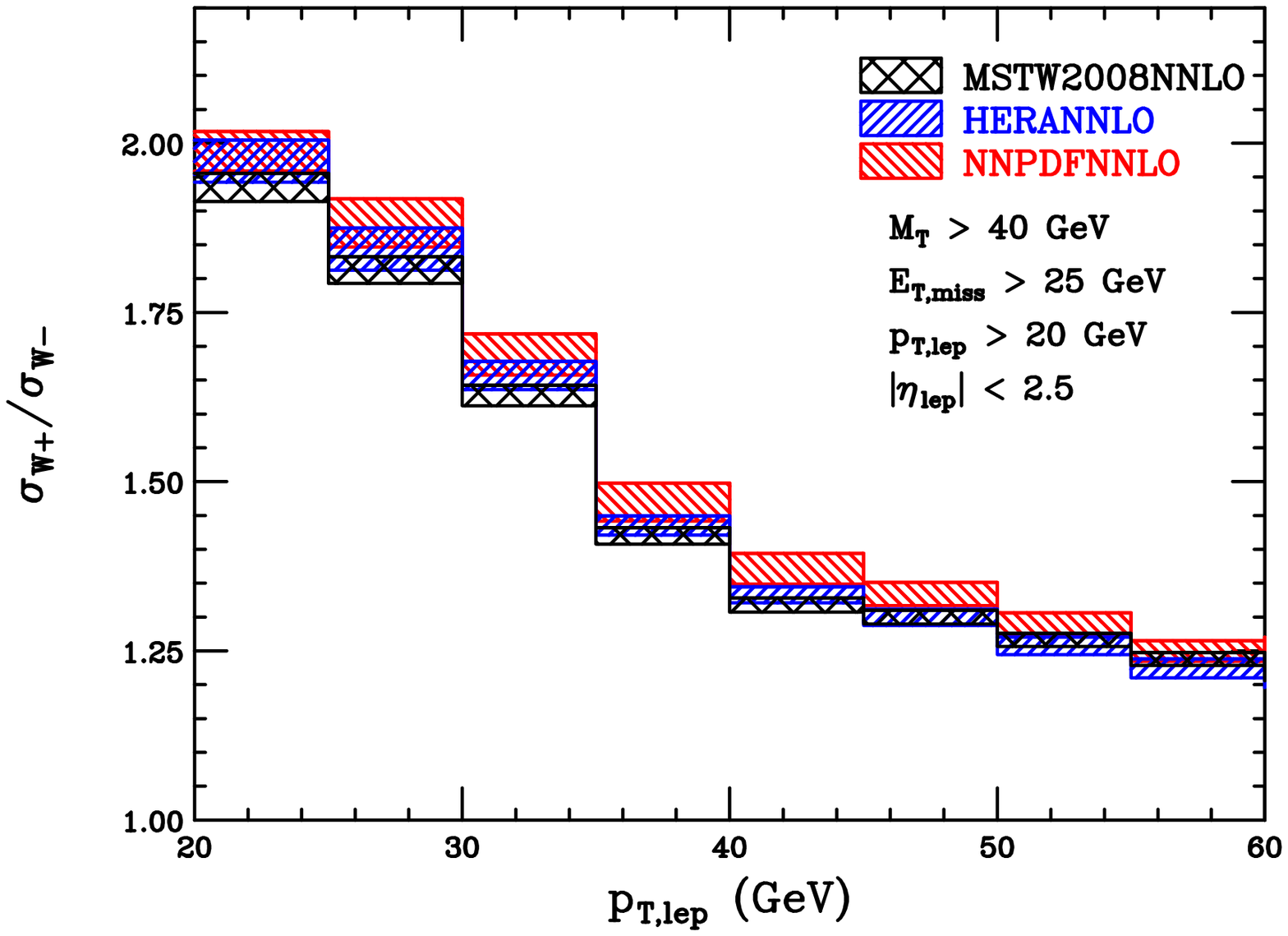}
\end{minipage}
\caption{Ratio of $W^+$ over $W^-$ production as a function of lepton
  pseudorapidity (left panel), and lepton $p_T$ (right panel).}
\label{ratios}
\end{figure}

\begin{figure}[h!]
\begin{center}
\includegraphics[width=4.0in]{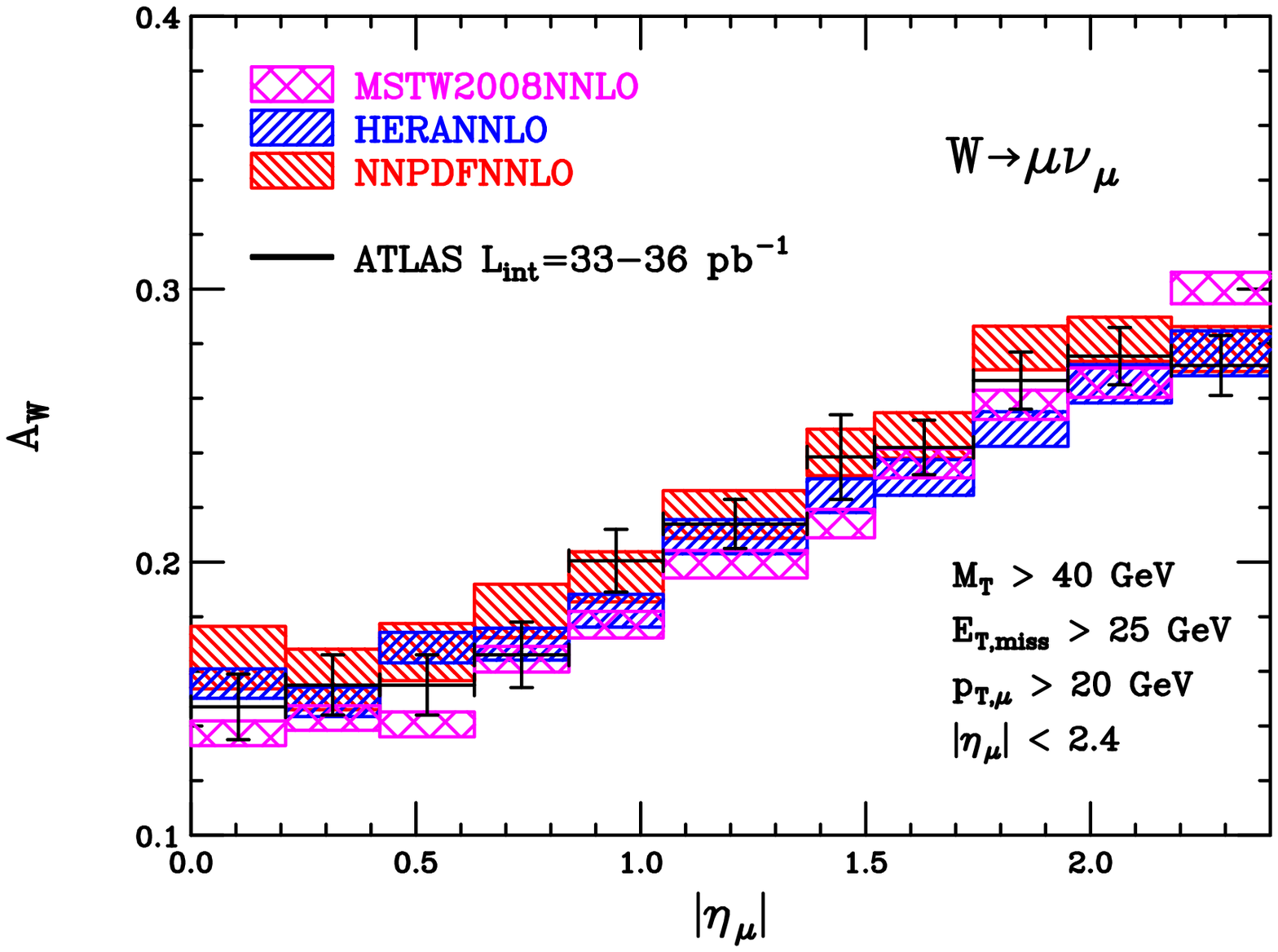}
\caption{Charge asymmetry at NNLO for three different PDF sets,
  compared with recent data from ATLAS.}
\label{charge}
\end{center}
\end{figure}

\section{Conclusions \label{sec:conc}}
We have presented an update to the program FEWZ, version 2.1, which
extends the functionality of FEWZ 2.0 to allow for the study of $W$
physics at hadron colliders.  Several new features have been added,
including new observables, the possibility of cumulative histograms or
ones with varying
bin sizes, and an interface to the PDF library LHAPDF.  Major speed
improvements have been achieved; NNLO precision in the presence of
standard acceptance cuts can now be achieved on a single machine in
hours.  We have presented example runs of the new code for $W$
observables with several NNLO PDF sets, both to show aspects of $W$
physics at the LHC and to demonstrate the array of studies possible
with the analysis scripts distributed with FEWZ 2.1. Comparisons to
early LHC data indicate excellent agreement of data and theory as
calculated by FEWZ.  With FEWZ, the extrapolation of measured
quantities to the full phase space and consequent increase in
experimental uncertainty can be avoided.  We look forward to the continued use of FEWZ in
understanding the properties of electroweak gauge bosons at hadron colliders with
unprecedented accuracy.

\bigskip \bigskip
\noindent
{\bf{\Large Acknowledgements}}

\medskip
\noindent
We thank U.~Klein for considerable input and testing during the initial
stages of integrating the $W$ code, and K. Mueller, W. Sakumoto, S. Stoynev, and
H. Yoo for substantial helpful feedback. This research is supported by
the US DOE under contract DE-AC02-06CH11357 and the grant
DE-FG02-91ER40684, and by the Swiss National Science Foundation.


\bigskip

\noindent

\begin{thebibliography}{99}

\bibitem{Hamberg:1990np} 
  R.~Hamberg, W.~L.~van Neerven and T.~Matsuura,
  Nucl.\ Phys.\ B {\bf 359}, 343 (1991)
  [Erratum-ibid.\ B {\bf 644}, 403 (2002)].


\bibitem{Anastasiou:2003yy}
  C.~Anastasiou, L.~J.~Dixon, K.~Melnikov and F.~Petriello,
  Phys.\ Rev.\ Lett.\  {\bf 91}, 182002 (2003)
  [hep-ph/0306192].

\bibitem{Anastasiou:2003ds}
  C.~Anastasiou, L.~J.~Dixon, K.~Melnikov and F.~Petriello,
  Phys.\ Rev.\  {\bf D69}, 094008 (2004)
  [hep-ph/0312266].

\bibitem{Melnikov:2006di}
  K.~Melnikov and F.~Petriello,
  Phys.\ Rev.\ Lett.\  {\bf 96}, 231803 (2006)
  [arXiv:hep-ph/0603182].
 
\bibitem{Melnikov:2006kv}
  K.~Melnikov and F.~Petriello,
  Phys.\ Rev.\  D {\bf 74}, 114017 (2006)
  [arXiv:hep-ph/0609070].
  
\bibitem{Catani:2009sm}
  S.~Catani, L.~Cieri, G.~Ferrera, D.~de Florian and M.~Grazzini,
  Phys.\ Rev.\ Lett.\  {\bf 103}, 082001 (2009)
  [arXiv:0903.2120 [hep-ph]].

\bibitem{Catani:2010en}
  S.~Catani, G.~Ferrera and M.~Grazzini,
  JHEP {\bf 1005}, 006 (2010)
  [arXiv:1002.3115 [hep-ph]].

\bibitem{Baur:1998kt} 
  U.~Baur, S.~Keller and D.~Wackeroth,
  Phys.\ Rev.\ D {\bf 59}, 013002 (1999)
  [hep-ph/9807417].

\bibitem{Dittmaier:2001ay} 
  S.~Dittmaier and M.~Kramer,
  Phys.\ Rev.\ D {\bf 65}, 073007 (2002)
  [hep-ph/0109062].

\bibitem{CarloniCalame:2006zq} 
  C.~M.~Carloni Calame, G.~Montagna, O.~Nicrosini and A.~Vicini,
  JHEP {\bf 0612}, 016 (2006)
  [hep-ph/0609170].

\bibitem{Cao:2004yy} 
  Q.~-H.~Cao and C.~P.~Yuan,
  Phys.\ Rev.\ Lett.\  {\bf 93}, 042001 (2004)
  [hep-ph/0401026].

\bibitem{Balossini:2009sa} 
  G.~Balossini, G.~Montagna, C.~M.~Carloni Calame, M.~Moretti, O.~Nicrosini, F.~Piccinini, M.~Treccani and A.~Vicini,
  JHEP {\bf 1001}, 013 (2010)
  [arXiv:0907.0276 [hep-ph]].

\bibitem{Bernaciak:2012hj} 
  C.~Bernaciak and D.~Wackeroth,
  arXiv:1201.4804 [hep-ph].

\bibitem{Kilgore:2011pa} 
  W.~B.~Kilgore and C.~Sturm,
  arXiv:1107.4798 [hep-ph].

\bibitem{Boughezal:2011jf} 
  R.~Boughezal, K.~Melnikov and F.~Petriello,
  arXiv:1111.7041 [hep-ph].

\bibitem{arXiv:1011.3540} 
  R.~Gavin, Y.~Li, F.~Petriello and S.~Quackenbush,
  Comput.\ Phys.\ Commun.\ \ {\bf 182}, 2388  (2011)
  [arXiv:1011.3540 [hep-ph]].

\bibitem{Whalley:2005nh}
  M.~R.~Whalley, D.~Bourilkov and R.~C.~Group,
  arXiv:hep-ph/0508110.

\bibitem{arXiv:0910.0467} 
  I.~W.~Stewart, F.~J.~Tackmann and W.~J.~Waalewijn,
  Phys.\ Rev.\ D\ {\bf 81}, 094035  (2010)
  [arXiv:0910.0467 [hep-ph]].
  
\bibitem{arXiv:1005.4060} 
  I.~W.~Stewart, F.~J.~Tackmann and W.~J.~Waalewijn,
  Phys.\ Rev.\ Lett.\ \ {\bf 106}, 032001  (2011)
  [arXiv:1005.4060 [hep-ph]].
  
\bibitem{hep-ph/0311311} 
  C.~Anastasiou, K.~Melnikov and F.~Petriello,
  Phys.\ Rev.\ D\ {\bf 69}, 076010  (2004)
  [hep-ph/0311311].
  
\bibitem{hep-ph/0501130} 
  C.~Anastasiou, K.~Melnikov and F.~Petriello,
  Nucl.\ Phys.\ B\ {\bf 724}, 197  (2005)
  [hep-ph/0501130].

\bibitem{Anastasiou:2010pw} 
  C.~Anastasiou, F.~Herzog and A.~Lazopoulos,
  JHEP {\bf 1103}, 038 (2011)
  [arXiv:1011.4867 [hep-ph]].

\bibitem{CooperSarkar:2011aa} 
  A.~M.~Cooper-Sarkar [ZEUS and H1 Collaborations],
  arXiv:1112.2107 [hep-ph].

\bibitem{Martin:2009iq} 
  A.~D.~Martin, W.~J.~Stirling, R.~S.~Thorne and G.~Watt,
  Eur.\ Phys.\ J.\ C {\bf 63}, 189 (2009)
  [arXiv:0901.0002 [hep-ph]].

\bibitem{Ball:2011uy} 
  R.~D.~Ball {\it et al.}  [The NNPDF Collaboration],
  Nucl.\ Phys.\ B {\bf 855}, 153 (2012)
  [arXiv:1107.2652 [hep-ph]].

\bibitem{Alekhin:2009vn}
  S.~Alekhin, J.~Blumlein, S.~Klein and S.~Moch,
  arXiv:0908.3128 [hep-ph].
  
\bibitem{JimenezDelgado:2009tv}
  P.~Jimenez-Delgado and E.~Reya,
  Phys.\ Rev.\  D {\bf 80}, 114011 (2009)
  [arXiv:0909.1711 [hep-ph]].

\bibitem{Aad:2011dm} 
  G.~Aad {\it et al.}  [ATLAS Collaboration],
  arXiv:1109.5141 [hep-ex].

\bibitem{CMS:2011aa} 
  S.~Chatrchyan {\it et al.}  [CMS Collaboration],
  JHEP {\bf 1110}, 132 (2011)
  [arXiv:1107.4789 [hep-ex]].

\bibitem{Watt:2011kp} 
  G.~Watt,
  JHEP {\bf 1109}, 069 (2011)
  [arXiv:1106.5788 [hep-ph]].

\bibitem{Berger:2010xi} 
  C.~F.~Berger, C.~Marcantonini, I.~W.~Stewart, F.~J.~Tackmann and W.~J.~Waalewijn,
  JHEP {\bf 1104}, 092 (2011)
  [arXiv:1012.4480 [hep-ph]].

\end{thebibliography}
\end{document}